\documentstyle[aps,prl,epsfig,twocolumn]{revtex}
\begin{document}
\twocolumn[\hsize\textwidth\columnwidth\hsize\csname
  @twocolumnfalse\endcsname
\preprint{IMSc-99/12/31}
\title{Spontaneous Time Asymmetry due to Horizon} 
\author{Ramesh Anishetty, R.Parthasarathy \cite{email}}
\address{The Institute of Mathematical Sciences, Chennai - 600 113, INDIA}
\maketitle
\begin{abstract}
We show that quantized matter fields in the presence of background metrics with
Horizon exhibit spontaneous time asymmetry. All quantized matter fields have to vanish
at the horizon. 
Some phenemenological applications 
of this in the context of black holes and early universe are
considered.

\end{abstract}
\vskip 2pc] 

Matter interaction in Minkowski space-time are well described by local quantum
field theories. These theories have Poinc\'{a}re invariance along with several
other continuous and discrete symmetries. We draw our attention to 
a discrete
symmetry namely time reversal symmetry $T$. This says
that the dynamics of the matter fields is such that if we evolve the system
backwards in time it is perfectly allowed, i.e. time symmetric. 
More precisely  the correlation
functions, for instance $<\phi(t)\phi(t')>$ as defined  by the Feynman's path
integral, are symmetric functions of $t-t'$ [1]. 
While this is self evident from the Lagrangian of the fundamental theory
, there are many physical situations which do not
appear to obey this symmetry. Some  well known examples [2] are given below. 
Decelerating electromagnetic charges radiate photons while the time reversed
phenomenon, namely,the accelerating charges absorbing photons, is never allowed.
Entropy in sufficiently large subspaces is always increasing even if the large
subspaces are causally disconnected. In the context of cosmolgy, we are
witnessing uniform expansion of the universe and growth of inhomogenieties
thereof and never observed the time reversed phenomenon so far.
These examples are suggesting that the vaccum state of the universe does seem
to break Time reversal invariance and thereby setting the arrow of time. 

In the Standard Model of particle physics [3],  
there is a small parameter which violates $CP$ invariance
 where
$C$ stands for charge conjugation and $P$ for spatial parity. Since CPT is still a good
symmetry[4], it implies $T$ has to be broken. The origin of this parameter in the
theory is not understood. 

It is also true that Time asymmetry can be
arranged by appropriate boundary conditions in time [5].
This may model the
phenomenon but it still begs the question: why that particular boundary
condition? In this paper we suggest a mechanism which imposes a
boundary condition due to dynamics itself and thus exhibit 
{\it{spontaneous Time
asymmetry}}.

We now turn to gravity and its coupling to matter as described by Einstein's 
equation. The quantum theory of gravity is a major unsolved problem of the
century. Inspite of this, some limited progress is made in realms where we
suspect the back recation of matter fluctuations on gravity is small due to 
small gravitational coupling constant[6]. In this realm gravity does influence
classically as background gravity. Indeed our understanding of cosmolgy after
the Planck's time rests on this assumption.

We address Time reversal invariance of matter fields in the presence of
background gravity. The metric is Minkowskian at any space
time point; however globally
it is  not Minkowskian.A generic background metric does not have any 
symmetries.Phenomenologically [7]the background gravity of the
universe does appear to have some symmtries: how this comes about is an
another interesting issue. Here we will assume that the background
gravity does have time translational invariance. That is, there is
a time like
Killing vector. 
By a suitable general co-ordinate transformation we can always choose the
metric  as given by the proper distance $ds$ 
to be
\begin{eqnarray}
ds^2&=&\sqrt{B(\vec{x})}\ (A(\vec{x})dt^2-g'_{ij}(\vec{x}) 
dx^idx^j),
\end{eqnarray} 
where $A,B$ and $g'_{ij}$ the three dimensional metric  
are all time independent. We further made 
a choice of definition such that the four volume density
$\sqrt{-det g}\ =\ B$ and $A\ det\ g'\ =\ 1$. $B$ field is
necessarily taken to be nonzero and positive everywhere. (If
$B$ is negative then it corresponds to wrong signature metric
and if it is zero, then it corresponds to dimensional
reduction of space time. These alternatives can be entertained
in some formalisms of gravity; we shall not consider them
here.)  
If the
field $A$ is positive everywhere then the time translation operator   $\frac{d}{dt}$ is time-like
, and the dynamics on such metrics have the same qualitative features as
that of  Minkowski space.

Consider a situation where $A$ becomes negative in some region of space.
In that region, the Killing vector, namely, the translation operator in $t$, will be space-like.  
By continuity, $A$ then has to vanish on some region. 
The region where $A(\vec{x})=0$ defines a
two-surface when the space dimension is three. This two-surface is what we
call the Horizon. Horizon in general can be open or closed and if closed it
could be topologically non-trivial as well. 

In the literature some of the well
known examples assume spherical symmetry. A convenient choice
of the metric which is static and spherically symmetric is [7]
\begin{eqnarray}
ds^2 &=& B(r)[A(r)dt^2 - A^{-1}dr^2] - r^2\ d{\Omega}^2,
\end{eqnarray}
with $\sqrt{-det g}\ =\ Br^2sin\theta$. The following examples
assume sphere topology for the Horizon : for static de Sitter
space, $B(r)\ =\ 1$ and   
\begin{eqnarray}
A(r)&=&1-\frac{r^2}{\alpha ^2},
\end{eqnarray}
where $\alpha$ is a constant; static
black hole solutions which approach Minkowski space time at 
infinity: $B(r)\ =\ 1$ and   
\begin{eqnarray}
A(r)&=& 1-\frac{2M}{r}+\frac{C^2}{r^2},
\end{eqnarray}
where $M$ is related to the mass and $C^2$ is to the charge of the black hole. Near the Horizon, the Ricci tensor is not singular.
(The apparent singularity in the metric at $r\ =\ 2M$ for $C\
=\ 0$ in (4), can be removed by Eddington-Finkelstein  
transformation and this introduces $drdt$ term in (2).
Based on this, time asymmetry had been argued in [8].)   
It is
however conjectured based on the positive energy condition that almost all
Riemann curvature singularities are hidden behind the Horizon.
(cosmic censorship hypothesis)[9].  
For example in (4), for $C=0$, the Riemann
curvature is singular as $M/r^3$ near $r=0$ which is enclosed by the horizon which
is a sphere of radius $2M$. There are other situations such as static de
Sitter (3) or
Rindler spaces  where the horizon is not enclosing singularities. It is noticed
that in these other examples the space at infinity is not Minkowskian.

Now we shall analyze how horizon generates {\it{spontaneous}} time reversal
non-invariance. The essential ingredients of our proof are already contained in
the spherically symmetric context. As for matter fields, for technical
simplicity, we consider a scalar field of mass $m$ in a spherical symmetric 
background metric (2), given by the action
\begin{eqnarray}
S&=&\frac{1}{2} \int dt dr r^2 d\Omega \Bigl\{A^{-1}{\bigl( 
\partial_{t}\Phi \bigr)}^2-A{\bigl(\partial_{r}\Phi  
\bigr)}^2-B\Bigl[\frac{1}{r^2}{\bigl(\partial_{\theta}\Phi \bigr)}^2
\nonumber \\
&+&\frac{1}{r^2 sin^2\theta}{\bigl(\partial_{\phi}\Phi \bigr)}^2
+m^2{\Phi}^2\Bigr] \Bigr\}.
\end{eqnarray}

The quantum theory can be envisaged by the following path integral given  by
the partition function $Z$, 
\begin{eqnarray}
Z&=&\int [d\Phi] exp{(iS)}.
\end{eqnarray}
The Lagrangian density in the above is manifestly invariant under time
reversal $t\rightarrow  -t$ 
with $\Phi$ transforming as a scalar.A more precise definition 
of the functional integral is given by expanding the field 
$\Phi $ into 
the eigen modes $\Psi $ 
of the covariance operator, namely of the eigenvalue problem
\begin{eqnarray}
\frac{1}{B}\Bigl\{-A^{-1}\partial_t^2+\frac{1}{r^2}{\partial}_r\ r^2A
{\partial}_r+B\Bigl[\frac{1}{r^2sin\theta}
{\partial}_{\theta}(sin\theta {\partial}_{\theta}) \nonumber
\\ 
+\frac{1}{r^2sin^2\theta}{\partial}^2_{\phi}   
-m^2\Bigr]\Bigr\}\Psi\ =\ \Lambda \Psi, 
\end{eqnarray} 
where $\Lambda $ are the eigenvlues and $\Psi $ the eigen modes.

Due to translational invariance in $t$, a good choice of eigen modes are
\begin{eqnarray}
\Psi &=& exp(-iEt){\psi}_E(r)Y_{jm}(\theta,\phi),
\end{eqnarray}
where ${\psi}_E$ is the radial and $Y_{jm}$ are the angular momentum parts.The radial
function satisfies the following equation on mass shell i.e. when 
$\Lambda \ =\ 0$. 
\begin{eqnarray}
&\frac{1}{B}\Bigl\{E^2A^{-1}+\frac{1}{r^2}{\partial}_r r^2A{\partial}_r
-B[\frac{j(j+1)}{r^2}&  \nonumber \\ 
&+m^2]\Bigr\}{\psi}_E(r)=0. & 
\end{eqnarray}

In the succeeding analysis we would like to assert some general features of the 
eigenvalues $E$. Let us say that the horizon is at 
$r=r_{+}$. $A(r)$ for $r\ >\ r_{+}$
is positve and negative for $r\ <\ r_{+}$. Near $r_{+}$
\begin{eqnarray} 
A(r)&\simeq & a(r-r_{+}).
\end{eqnarray}
Usually [6,10], the quantum theory is  studied outside the
horizon. We shall consider this region first. 
Consider any state which is nonvanishing in the region $r \geq r_{+}$.
The first term in the  eqn.(9) has a pole at $r = r_{+}$, this can be
cancelled only if the eigenfunction near $r_{+}$ has the behaviour
\begin{eqnarray}
\psi_{E}(r)&\simeq &K(r-r_{+})^{iE/a}.
\end{eqnarray} 
Therefore the eigenfunction oscillates wildly near $r=r_{+}$ if $E$ is positive or negative. In general we
can allow $E$ to be complex. However if imginary part of $E$ is  positive the
eigenfunction
blows up near the horizon and hence not normalizable. If imaginary part of 
$E$  is negative
the eigenfunction necessarily vanishes at the horizon. We will  show that
only the latter is the allowed alternative.
Taking scalar product with (9) i.e. multiplying with $Br^2dr \psi^*_E(r)$      
and integrating from $r_{+}$ to infinity, we obtain
\begin{eqnarray}
&E^2\int_{r_{+}}dr r^2 A^{-1}{\mid
\psi_{E}\mid}^2+\int_{r_{+}}dr \psi^*_{E}(r){\partial}_r
(r^2A{\partial}_r\psi_{E}(r))& \nonumber \\
&-\int_{r_{+}}B[j(j+1)+m^2r^2]{\mid\psi_E(r)\mid}^2\ =\ 0.& 
\end{eqnarray}

Performing the integration by parts in the second term, we
obtain
\begin{eqnarray}
&E^2\ =\ I\Bigl\{
\int_{r_{+}}dr \Bigl(r^2A{\mid{\partial}_r\psi_{E}\mid }^2
+B[j(j+1)  
+m^2r^2]& \nonumber \\ 
&{\mid\psi_{E}\mid }^2\Bigr)  
+iK^2 Er^2_{+}({Limit}_{r\rightarrow r_{+}})
(r-r_{+})
^{\frac{i}{a}(E-E^*)}\Bigr\}, &  
\end{eqnarray}
where $I^{-1}\ =\ \int dr r^2 A^{-1}{\mid \psi_E\mid }^2$. 
We have assumed that the eigenfunction and its derivative vanish for large $r$. In
the region of integration $A$ is always positive. 

>From (13) we see that if the imaginary part of $E$ is
negative then it is infact zero and the real part of $E$ can be positive or 
negative. Moreover for every possible positve real $E$ there is also an eigenstate
with $-E$. If the imaginary part of $E$ is positive then 
it is also infinite and the radial integrals in (12,13) are also
not
normalizable.
Hence we conclude the allowed eigenvalues are
\begin{eqnarray}
E\ =\ {\cal{E}} -i \epsilon,
\end{eqnarray}
where $\cal{E}$ is the real part and $\epsilon$ is infinitesimally positive.

To summarize, the operator as defined by the left hand side of
(9) has 
normalizable eigenfunctions
only if imaginary part of $E$ is negative which in turn
implies the eigenfunction necessarily has to vanish at the horizon and all the
energy eigenvalues are real on shell. This shows any eigenfunction which is nonvanishing
outside the horizon, necessarily vanishes at the horizon. The
results (11) and (14) hold true for off-shell ($\Lambda \ \neq
\ 0$) case of (7) as well.  

Exactly the same analysis as above goes through inside the Horizon i.e for 
$r \leq r_{+}$. We again arrive at the conclusion E has only negative imaginary
part. Therefore we have shown that all eigenfunctions have a node at the
Horizon which inturn implies (14).

Consider the correlation function
\begin{eqnarray}
<\Phi(x)\Phi(x')>\ =\ \int dE d\omega
\frac{\Psi^*(x)\Psi(x')}{\Lambda
{(\cal{E}} -i \epsilon,\omega)},
\end{eqnarray} 
where $\Lambda$'s are the eigenvalues of the operator which in turn is explicitly 
 a function of $E \ =\ {\cal{E}}-i\epsilon $  
and other quantum numbers which we have collectively called
them
as $\omega$. The explicit form  is quite complicated in 
general. (For example see [10] for static de Sitter space
metric).  
The 
important feature is that now the correlation function is necessarily
time reversal non-invariant due to $\epsilon$. {\it{The imaginary part of $E$ has always only one
sign, this in turn implies that we only have retarded (or only advanced by the alternate
choice) propagation.  
This demonstrates that in spite of
manifest time reversal invariance of the Lagrangian the system picks a
time reversal non-invariant vacuum, i.e., the vacuum breaks time reversal
invarince spontaneously}}.

In contrast, in background Minkowski space, which has no horizon, we have many alternatives for
vacuum state, namely, only retarded or only advanced propagation or the average of the retarded
and advanced. The last is the choice made in the "Stuckelberg-Feynman $i\epsilon $ prescription",
[1,11] which is manifestly time reversal invariant. 
It is clear that the above argument actually rested totally on the zero of $A$.
Indeed all the above steps of the proof prevail even if we discard spherical
symmmetry and if there are more than one horizons which are disconnected.
This phenomenon is in fact quite generic whenever there are any horizons.

We can construct an example of an horizon which is open, such as the metric
given by (1) and $A(x,y,z)=x\ ;\ B\ =\ 1$ 
and $g_{xx}\ =\ 1/x;\ g_{yy}\ =\ g_{zz}\ =\ 1$ and all the
other components of the metric zero.  
The horizon is the (y,z) plane
Our previous analysis goes through exactly the same way leading to the same
conclusion that the energy eigenvalues are ${\cal{E}}-i
\epsilon$.
       
Matter self interactions do not alter the esssential conclusions. Matter
fields such as Fermionic or gauge fields also yield the same qualitative
result as discussed above. Namely they all generically exhibit a node
at the Horizon, however they can pass through it. 
Inkling of this phenomenon is already present in the
unquantized matter. For instance a particle moving along its geodesic in the
presence of Schwarzschild black hole, in its own reference frame, has zero
velocity and zero acceleration at the Horizon.

It is interesting to note that in all the cases the $A$ function has to vanish
linearily at the Horizon; if $A$ vanishes
quadratically or fractional power, then we cannot find any 
eigen solutions for (7). This pathology is resolved by noting that: 
A continuous function vanishing linearily is a generic 
situation. However when two horizons coalesce then the $A$ function does have
quadratic zeros and this situation is certainly not a static background; under
those circumstances our analysis does have to consider the interaction of
gravity and matter as a time dependent phenomenon which involves the Einstein
equation as well. We have not considered that here.

We consider some phenomenological applications of  this phenomenon,
namely horizons do cause time assymetry. Time asymetry is essential to
produce baryon number violation in the universe. We suggest that since
every black hole has a horizon they in turn might have the signature that
they produce more  particles
than anti-particles. Furthermore all quantized
matter fields have to have a node at the Horizon, this manifests itself as if
all particles
are repelled from the Horizon. This phenomenon can be 
used as a signature for compact black holes.  

In the context of early universe at Planck's time we need to consider both
matter and gravity quantum fluctuations. While this is least understood,
Feynman path integral for gravity and matter is an acceptable mathematical
framework. In this framework metric fluctuations are taken into account
with amplitude given by Einstein action or some modification thereof
which depend on the Riemann tensor. Let us consider metric fluctuations which
create Horizons. These configurations generically do not produce large
Riemann curvatures, infact near the Schwarzschild horizon Ricci tensor
vanishes. Hence as quantum fluctuations Horizons can be produced prolifically.
These Horizons need not necessarily be one of those which hide a singularity 
such as that of black holes. It is suggestive from functional integral that
the horizon as a quantum fluctuation can be one which has no singularity for the Ricci
tensor anywhere.
Our own early universe might have generated one such at the beginning.
This Horizon can cause the primodial Baryon anti-Baryon asymmetry.The
understanding of dynamics in these contexts where exchange of energy between
gravity and matter modes is profusive is yet to be done. We speculate that 
as matter field repel the Horizon, perhaps the early universe
Horizon manifests its presence by the present Hubble expansion of all matter
fields away from the center of the universe and thereby
setting the arrow of time. 
The creation of the Horizon
itself can be thought as the initial time of the universe. This creation being
on a two surface, the so called initial point singularity of the big bang [9] 
has no role to play.
This phenomenon 
warrants a more
careful study.

\end{document}